\documentstyle[epsfig,12pt]{article}

\newcommand{\bce}{\begin{center}} 
\newcommand{\ece}{\end{center}}
\newcommand{\beq}{\begin{equation}}
\newcommand{\eeq}{\end{equation}}
\newcommand{\bea}{\vspace{0.25cm}\begin{eqnarray}}
\newcommand{\eea}{\end{eqnarray}}

\newcommand{\brho}{\mbox{\boldmath $\rho$}}

\newcommand{\bk}{{\bf k}}

\newcommand{\br}{{\bf r}}

\newcommand{\ba}{\begin{array}}
\newcommand{\ea}{\end{array}}

\newcommand{\doublespace}{
    \renewcommand{\baselinestretch}{1.6}\large\normalsize}

\def\lsim{\mathrel{\rlap{\lower4pt\hbox{\hskip1pt$\sim$}}
    \raise1pt\hbox{$<$}}}         
\def\gsim{\mathrel{\rlap{\lower4pt\hbox{\hskip1pt$\sim$}}
    \raise1pt\hbox{$>$}}}         

\def\Pom{{\bf I\!P}}

\def\beq{\begin{equation}}
\def\endeq{\end{equation}}
\def\arr{\begin{eqnarray}}
\def\endarr{\end{eqnarray}}

\makeindex

  \advance\oddsidemargin  by -1in
  \advance\evensidemargin by -1in
  \advance\topmargin      by -0.5in
\begin{document}

\vspace{2.0cm}

\begin{flushright}
\end{flushright}

\vspace{1.0cm}

\begin{center}
{\Large \bf 
Color screening, absorption  and  $\sigma_{tot}^{pp}$ at LHC}\\

\vspace{0.5cm} 
{\large \bf R. Fiore$^{1}$,
N.N. Nikolaev$^{2}$ 
and  
V.R. Zoller$^3$}\\
\vspace{0.5cm}
$^1)${\em Dipartimento di Fisica,
Universit\`a     della Calabria\\
and\\
 Istituto Nazionale
di Fisica Nucleare, Gruppo collegato di Cosenza,\\
I-87036 Rende, Cosenza, Italy}

$^{2)}${\em L.D. Landau Institute for  Theoretical  Physics,\\
Chernogolovka 142432, 
Moscow Region, Russia} 

$^{3)}${\em Institute for  Theoretical and Experimental Physics,
Moscow 117218, Russia} 
\vspace{1.0cm}

{ \bf Abstract }\\
\end{center}

\vspace{1.0cm}
We show that a growth of the proton-proton total cross section with energy
 can be entirely
 attributed to the purely perturbative mechanism.  
The infrared regularization at rather short distances
 $R_c\simeq 0.3$ fm
allows to  extend the BFKL technique
from deep inelastic  to  hadron-hadron scattering. 
With the account of the absorption  corrections  our results  
are in  agreement with the LHC data on  $\sigma_{tot}^{pp}$.

\doublespace

\vskip 0.5cm \vfill $\begin{array}{ll}
\mbox{{\it email address:}} & \mbox{roberto.fiore@cs.infn.it} \\
\mbox{{\it email address:}} & \mbox{zoller@itep.ru} \\
\end{array}$

\pagebreak



{\bf 1. Introduction.}

 In  deep inelastic scattering (DIS) of leptons on nucleons, 
 the density of  BFKL 
\cite{BFKL} gluons has been established to 
 grow fast to smaller values of Bjorken $x$, $xg(x)\sim x^{-\Delta}$,
where, phenomenologically, $\Delta\simeq 0.3$. 
 In practice, the perturbative QCD base phenomenology of DIS structure functions 
is rather sensitive
 to the infrared regularization  which defines
 a transition between  the
 nonperturbative and perturbative domains.
It is generally accepted, that in the QCD vacuum the non-perturbative fields form
 structures with sizes $\sim R_c$ significantly smaller than 
$\Lambda^{-1}_{QCD}$ and local field strengths much larger 
than $\Lambda^{2}_{QCD}$. Instantons are one of prominent candidates \cite{Shuryak}.
A direct confirmation of this picture comes from the lattice 
\cite{DIGIACOMO}. 
The non-perturbative fluctuations in the QCD vacuum restrict the phase 
space for the 
perturbative (real and virtual) gluons.
 The perturbative gluons with short propagation length,
 $R_c\sim 0.2-0.3$ fermi, as it follows from the fits to lattice data 
on field strength correlators \cite{DIGIACOMO}, do not walk to large distances, 
$r >  R_c$. This is the vacuum  color screening effect. 

Explicit IR regularization with such a small  $ R_c$,
allows one to  extend the BFKL technique
from DIS to  hadron-hadron scattering.
Take for instance proton-proton scattering. There is always a contribution
from small-size dipoles in the proton to the  color dipole factorization
formula 
$\sigma^{pp}_{tot}=\int d^2r|\Psi_p(r)|^2\sigma(r).$
For example, in the symmetric oscillator approximation for the 3-quark proton, a probability $w_{p}(r<R_c)$
to find dipoles of
size $r\lsim R_c$ can be estimated as 
\beq
w_{p}(r<R_c)\simeq {R_c^2\over 2\langle r^2_{p}\rangle}.
\label{eq:WPROB}
\eeq
In this approximation the  proton looks as $3/2$
color dipoles spanned between quark pairs and
 $\langle r^2_{p}\rangle = 0.658\, {\rm fm^2}$ as
suggested by the standard dipole form factor of the proton.
The latter gives quite a substantial fraction of the proton,
\beq
w_{p}(r<R_c)\simeq 5\cdot 10^{-2}
\label{eq:WNUM}
\eeq
the interaction of which with the target nucleon proceeds in the
 hard regime typical of DIS. The corresponding contribution
to $\sigma_{tot}^{pp}$ must exhibit the same rapid rise with energy as
the proton structure function. Furthermore, in the BFKL approach
there is always a diffusion in the dipole size by which there is
a feedback from hard region to interaction of large dipoles and
vice versa. At large  $r\gsim R_c$
 a sort of the additive quark model is recovered:
the quark of the dipole $\vec r$ develops its own perturbative 
gluonic cloud and  gluonic clouds of different quarks 
 do not overlap at $ r\gg R_c$.

 Below we discuss  how substantial such a hard  BFKL
 contribution to the
proton-proton  total cross section could be.

 For high enough parton densities the 
phenomenon of 
 parton fusion  becomes important \cite{Kancheli73,NZ75}. Corresponding
 unitarity, e.g. absorption   corrections to the BFKL evolution are
 described by the non-linear BK  equation \cite{B,K}.
 The strength of non-linear effects 
 depends crucially on  the IR cutoff $R_c$
\cite{FSZ2012,FZPLB,FZSLOPE}. 
Hence, one more issue we address in this communication is 
the role of the absorption corrections to $\sigma^{pp}_{tot}$ and the non-linear
dynamics of the perturbative component of 
$pp$-interactions at the LHC energies. For alternative approaches to the problem of 
$\sigma_{tot}^{pp}$ at superhigh energies see \cite{Kostya,Ostap}.

{\bf 2. Vacuum color screening and CD BFKL}

 A distribution of perturbative gluons around the quark source is described by light 
cone radial wave function $\psi({\brho})$ 
\beq
\psi({\brho})={\sqrt{C_F\alpha_S(R_i)}\over \pi}{{\brho}\over \rho R_c}
K_{1}(\rho/R_c),
\label{eq:PSIQG}
\eeq
where the modified Bessel function,  $K_{1}(t)$, parameterizes the exponential decay of the 
perturbative gluon fields by vacuum screening at large distances,  $r > R_c$\cite{NZZJL94,NZJETP94}

The effects of finite  $R_c$ are consistently  incorporated 
 by the generalized color dipole (CD) BFKL equation 
(hereafter CD BFKL)\cite{NZZJL94,NZJETP94}.
\bea
{\partial_{ \xi} \sigma(\xi,r)}
=\int d^{2}{\brho}_{1}\,\,
\left|\psi({\brho}_{1})-\psi({\brho}_{2})\right|^{2}\nonumber\\
\times\left[\sigma_3(\xi,\br,\brho_1,\brho_{2})-\sigma(\xi,r)\right] ,
\label{eq:CDBFKL}
\eea
where  the 3-parton ($q\bar q g$-nucleon) cross section is
\beq
\sigma_3(\xi,\br,\brho_1,\brho_2)={C_A\over 2C_F}\left[\sigma(\xi,\rho_1)+
\sigma(\xi,\rho_{2})- \sigma(\xi,r)\right]+\sigma(\xi,r),
\label{eq:SIGMA3}
\eeq
where $C_A=N_c$ and $C_F=(N_c^2-1)/2N_c$.
Denoted by $\brho_{1,2}$ are the $q$-$g$ and $\bar{q}$-$g$ separations
in the two-dimensional impact parameter plane
for dipoles generated by the $\bar{q}$-$q$
color dipole source.
 The one-loop  QCD coupling 
\beq
\alpha_{S}(R_i)=4\pi/\beta_0\ln(C^2/\Lambda^2_{QCD}R_i^2)
\label{eq:ALPHAS}
\eeq
 is  taken at 
the shortest relevant
distance $R_{i}={\rm min}\{r,\rho_{i}\}$. 
In the numerical analysis $C=1.5$, $\Lambda_{QCD}=0.3$ GeV,
 $\beta_0=(11N_c-2N_f)/3$  and
 infrared freezing $\alpha_{S}(r>r_f)=\alpha_f=0.8$  has been imposed

 The BFKL dipole  cross section $\sigma(\xi,r)$, 
where $\xi=\ln(x_0/x)$ and $r$ is the $q\bar q$-separation,
sums the Leading-Log$(1/x)$ multi-gluon production cross
sections within the QCD perturbation theory (PT).
As a
realistic boundary condition for the
BFKL dynamics we take the lowest PT order $q\bar q$-nucleon cross section at 
some $x=x_0$. It is described by   the  Yukawa screened 
two-gluon exchange and is basically parameter-free one.

{\bf 3.  Non-perturbative component of the dipole cross section.}

The perturbative gluons are confined and 
do not propagate to large distances. Available fits \cite{DIGIACOMO} to the lattice QCD 
data suggest Yukawa screening of perturbative color fields with 
propagation/screening 
radius $R_c\approx 0.2-0.3$ fm. The value $R_c=0.275$ fm has been used since 1994 
in the very successful color dipole phenomenology of small-x DIS 
\cite{NZHERA,NZZEXP,NZZ97,SlopeJETP98,CharmBeauty}. 
Because the propagation radius is short compared to the typical range of strong 
interactions the dipole cross section obtained as a solution of the CD BFKL
 equation
(\ref{eq:CDBFKL}) would miss the interaction strength for large color dipoles. 
In \cite{NZHERA,NZZEXP} this missing strength was modeled by the $x$-independent
 dipole cross section, so that our heterotic solution is that he perturbative, 
$\sigma(\xi,r)$,  and non-perturbative, $\sigma_{npt}(r)$, cross sections 
are additive,
\beq
\sigma_{tot}(\xi,r)=\sigma(\xi,r)+\sigma_{npt}(r).
\label{eq:PTNPT}
\eeq

The principal point about the non-perturbative component of $\sigma_{tot}(\xi,r)$
is that it must not be subjected to the perturbative BFKL evolution. 
Thus, the 
arguments about the rise of $\sigma(\xi,r)$ due to the hard-to-soft diffusion
 do not apply to  $\sigma_{npt}(r)$. 
We reiterate, finite $R_c$ means that gluons with
the wave length $\lambda \gsim R_c$ are beyond  the realm of  
perturbative QCD.  Therefore,  the intrusion of hard regime into soft $pp$-scattering
 is the sole source
of the rise of total cross sections.  
Specific form of $\sigma_{npt}(r)$ used in the present paper is found in \cite{FZPLB}.

{\bf 4. Non-linear regime. Absorption effects..}
 
 We considered above  the non-unitarized running CD BFKL amplitudes 
too rapid a rise of which
must be tamed by the unitarity absorption corrections.
The simplest way to take them into account was suggested first in \cite{B,K}.
In \cite{FZPLB} the BK equation was  rederived
 in terms of 
 the $q\bar q$-nucleon  partial-wave amplitudes 
(profile functions) and 
 for the  predominantly imaginary
 elastic dipole-nucleon amplitude $f(\xi,r,\bk)=i\sigma(\xi,r)\exp(-Bk^2/2)$
upon integrating over the impact parameters it was reduced to the following form 
 \cite{FZPLB}
\bea
{\partial_{\xi}\sigma(\xi,r)} =
 \int d^{2}{\brho}_{1}\,\,
\left|\psi({\brho}_{1})-\psi({\brho}_{2})\right|^{2}\nonumber\\
\times\left\{\sigma(\xi,\rho_{1})+
\sigma(\xi,\rho_{2})-\sigma(\xi,r)\right.\nonumber\\
\left.-{\sigma(\xi,\rho_{1})\sigma(\xi,\rho_{2})\over 4\pi(B_1+B_2)}
\exp\left[-{r^2\over 8(B_1+B_2)}\right]\right\},
\label{eq:BFKLNL}
\eea
where $B_i=B(\xi,\rho_i)$.
This form of equation with the above  definition of the elastic amplitude  $f$
  removes uncertainties with the radius $R$ of the area within which 
 interacting gluons  are expected to be distributed, thus removing the 
frequently used in the literature parameter $S_{\perp}=\pi R^2$.
The diffraction slope 
 for the forward cone in the dipole-nucleon scattering is \cite{SL94,SLPL}
\beq
B(\xi,r)= {1\over 2}\langle {\bf b}^2\rangle= {1\over 8}r^2+{1\over 3} R_N^2 
+2\alpha^{\prime}_{\Pom}\xi,
\label{eq:BSLOPE}
\eeq
where
 $r^2/8$ is the purely geometrical term related to the elastic 
form factor of the 
color dipole of the size $r$, $R_N$ represents
the gluon-probed radius of the proton, 
the dynamical 
component of $B$ is given by the last term in Eq. (\ref{eq:BSLOPE}) 
 where 
$\alpha^{\prime}_{\Pom}$ is the Pomeron trajectory slope evaluated first 
in \cite{SL94} (see also \cite{SLPL}). 
Here we only cite the order of magnitude estimate \cite{SLPL} 
\beq
\alpha^{\prime}_{\Pom}\sim
{3 \over 16\pi^{2}} \int d^{2}\vec{r}\,\,\alpha_{S}(r)
R_c^{-2}r^{2}
K_{1}^{2}(r/R_c)  \sim
{3 \over 16\pi}\alpha_{S}(R_{c})R_{c}^{2}
\, ,
\label{eq:ALPRIME}
\endeq
which  clearly shows the connection between the dimensionful
$\alpha^{\prime}_{\Pom}$ and the non-perturbative infrared parameter
$R_{c}$. 

In Eq. (\ref{eq:BSLOPE}) the gluon-probed radius of the proton is a 
phenomenological parameter to be determined from the experiment.
The analysis of Ref. \cite{INS2006} gives
$R_N^2 \approx 12{\, \rm GeV}^{-2}$.

{\bf 5. Absorption and  large dipoles, $r\gsim R_c$.}

The proton size $r_p$ is much larger than the correlation radius $R_c$. 
 In high-energy scattering
 of  large  dipoles, $r\gg R_c$,
 a sort of the additive quark model is recovered:
the quark of the dipole $r$ develops its own perturbative 
gluonic cloud and the pattern of diffusion changes dramatically. Indeed,
in this region the term proportional to 
$K_1(\rho_1/R_c)K_1(\rho_2/R_c)$ in the 
kernel of Eq. (\ref{eq:CDBFKL}) is exponentially small, what is related 
to the exponential decay
of the correlation function (the propagator) of perturbative gluons. 
Then, at large $r$ the kernel  will be dominated by the 
contributions from $\rho_1\lsim R_c\ll\rho_2\simeq r$ and from
$\rho_2\lsim R_c\ll\rho_1\simeq r$. It does not depend on $r$ and 
 for large $N_c$ the equation for the dipole cross section reads
\bea
{\partial_{\xi}\sigma(\xi,r)}={\alpha_S C_F\over \pi^2}\int
 d^2\rho_1  R_c^{-2}K_1^2(\rho_1/R_c)\nonumber\\
\left\{\sigma(\xi,\rho_1)+\sigma(\xi,\rho_2)
-\sigma(\xi,r)\right.\nonumber\\
\left.-{\sigma(\xi,\rho_1)\sigma(\xi,\rho_2)\over 4\pi(B_1+B_2)}
\exp\left[-{r^2\over 8(B_1+B_2)}\right]\right\},
\label{eq:LARGE1}
\eea 
where $B_i=B(\xi,\rho_i)$. 
 
 From   Ref.\cite{FZPLB} it follows that the absorption 
correction to the dipole cross section is 
\bea
\delta\sigma \sim R_c^{-2}\int^{R_c^2} {d\rho^2}K_1^2(\rho/R_c)
{\sigma(\xi,\rho)\sigma(\xi,r)\over 8\pi B}\sim\nonumber\\
\sim{\sigma(\xi,R_c)\sigma(\xi,r)\over 8\pi B}.
\label{eq:deltaSigma}
\eea
With growing $\xi$ the dipole cross section $\sigma(\xi,r)$  
increases approaching the unitarity bound, $\sigma=8\pi B$.
Untill ${\sigma(\xi,R_c)/ 8\pi B}\ll 1$
\beq
{\delta\sigma_{tot}\over\sigma_{tot}}\sim{\sigma(\xi,R_c)\over8\pi B}.
\label{eq:kappa}
\eeq
The Eq.~(\ref{eq:kappa}) explains why the absorption correction 
$\delta\sigma_{tot}$
dominated by the dipoles of  sizes $r\sim R_c$
grows with energy faster than $\sigma_{tot}$ dominated by $r\sim r_p$.
The point is that  the local pre-asymptotic pomeron intercept
$\Delta(\xi,r)$ in the parameterization
\beq
\sigma(\xi,r)\propto \exp[\Delta(\xi,r)\xi]
\label{eq:EXPDELTA}
\eeq
depends on $r$ \cite{NZZJETP94} and for  $R_c \ll r_p$
 \cite{NZZJETP94}
\beq 
\Delta(\xi,R_c)>\Delta(\xi,r_p)
\label{eq:DEL1DEL2}
\eeq
\begin{figure}[h]
\psfig{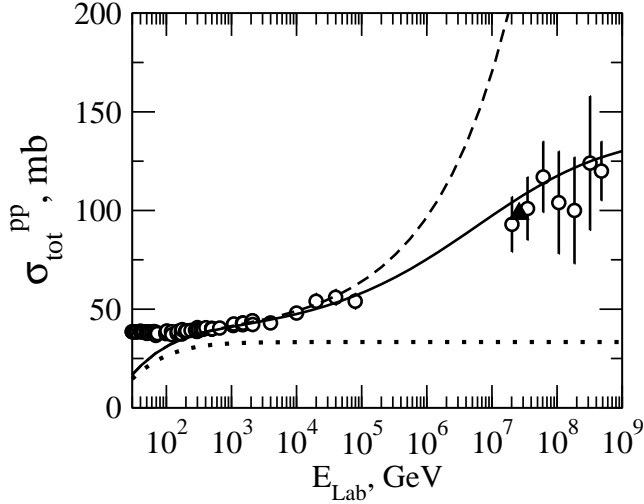}
\vspace{-0.5cm}
\caption{The CD BFKL description 
 of the experimental data \cite{PDG2013} on $\sigma^{pp}_{tot}$. 
Dashed line corresponds to  $\sigma^{pp}_{tot}$ obtained within 
the linear CD BFKL with color screening. The account of the
 absorption corrections results in $\sigma^{pp}_{tot}$ shown by the
solid line.
Doted line - the nonperturbative contribution to $\sigma^{pp}_{tot}$.
 The  black triangle corresponds to  $\sigma^{pp}_{tot}$ as 
 measured by  the LHC
 \cite{PDG2013}.}
\label{fig:fig1}
\end{figure}  

{\bf 6. Comparison with experimental data .}

In  \cite{FZPLB} we found that
the  choice $R_{c}=0.26$\,fm  leads to a very good
description  of the DIS  data
 on the proton structure
function  $F_2(x,Q^2)$ at small $x$.
Applying the color dipole factorization to $\sigma^{pp}_{tot}$ 
 we observe (see Fig.~1) that hard effects in the
 $p p$ scattering 
 do exhaust  completely
the observed rise of $\sigma_{tot}^{pp}(E_{Lab})$ at moderately large $E_{Lab}$.
However, if the CD BFKL evolution is treated to a linear approximation, then 
the predicted  $\sigma_{tot}^{pp}(E_{Lab})$ 
would exhibit too rapid a rise at superhigh energies. 
The real issue is whether there exists a mechanism 
to tame this excessive growth of $\sigma_{tot}^{pp}(E_{Lab})$ at very high-energies.
We addressed this issue resorting to the BK-equation \cite{B,K} reformulated  to incorporate
the effects of the finite correlation length of perturbative gluons (see Eq.\ref{eq:BFKLNL}).
 Shown by the solid line in Fig.~1 is the $pp$ total cross section 
evaluated with the account of the absorption effects.
The agreement with data is quite reasonable. In Eq.(\ref{eq:BFKLNL})
The plausible choice of the Regge parameter is
${1/ x}={W^2/ M^2}\,,$ with $W^2\approx 2m_p E_{Lab}$
and $M^2=1.5$ GeV$^2$

The emerging hierarchy of perturbative
and nonperturbative components of the $pp$ total cross section is as  follows. At moderately high
energies, $E_{Lab} < 10^4-10^5$ GeV, the  bulk of the total cross section comes 
from the nonperturbative large dipoles. The purely perturbative component of the total cross section
is still the subdominant one, is capable of describing a growth
of the total cross section, and receives only marginal  nonlinear corrections. 
At superhigh energies, $E_{Lab} > 10^4-10^5$ GeV, the perturbative component starts taking over
and its unitarization becomes a central issue.
In \cite{FSZ2012,FZPLB,FZSLOPE} it was demonstrated that the non-linear, i.e., unitarity effects
 are very sensitive to the IR regularization which defines a transition between  the
 nonperturbative and perturbative domains. Empirically, the  bulk of the total cross
 section at  $E_{Lab}\sim 10^4-10^5$ GeV comes from the nonperturbative domain and our analysis
 suggests rather small, infrared cutoff for the perturbative contribution, $R_c=0.26$ fm. 
In an alternative approach to the purely perturbative non-linear analysis,  Ref.~\cite{AAA}
suggests  a rather soft 
infrared regularization at distances 
$\approx 0.8$ fm.  Normally,
 the QCD perturbation theory would completely  break down, and the purely
 perturbative BFKL and BK analyses would be nonsensical, at such a large
 distances. The authors \cite{AAA}
circumvent
 the problem by enforcing a surprisingly small  running  
QCD coupling
which is $\alpha_S\approx 0.45$ at  $r=0.8$ fm.
 
\section{Conclusions}

We explored the consequences for 
 high energy total cross sections from the BFKL dynamics with finite
 correlation
 length of perturbative gluons, $R_c$. We
use very restrictive  perturbative two-gluon exchange as a 
parameter-free boundary condition for  BFKL and  BK evolution in the 
color dipole basis. 
 With the account of the  BK absorption
 and under plausible assertions on the color dipole 
structure of the proton, our parameter-free description of
 the total proton-proton cross section  agrees reasonably 
well with the LHC determinations.

{\bf Acknowledgments.} 
V.R.~Z. thanks
 the Dipartimento di Fisica dell'Universit\`a
della Calabria and the Istituto Nazionale di Fisica
Nucleare - gruppo collegato di Cosenza for their warm
hospitality while a part of this work was done.
The work was supported in part by the Ministero Italiano
dell'Istruzione, dell'Universit\`a e della Ricerca,   by
 the RFBR grants 11-02-00441 and 12-02-00193.

\end{document}